\documentclass[aps, prd, twocolumn, superscriptaddress, nofootinbib]{revtex4-1}
\usepackage{graphicx}
\usepackage{dcolumn}
\usepackage{amssymb}

\begin{document}

\newcommand{\BHKU}{BHKU}
\newcommand{\fd}{\ensuremath{f_{10}}}
\newcommand{\ns}{\ensuremath{n_{\mathrm{s}}}}
\newcommand{\Ob}{\ensuremath{\Omega_{\mathrm b}}}
\newcommand{\Oc}{\ensuremath{\Omega_{\mathrm c}}}
\newcommand{\Om}{\ensuremath{\Omega_{\mathrm m}}}
\newcommand{\Obhh}{\ensuremath{\Ob h^{2}}}
\newcommand{\Omhh}{\ensuremath{\Om h^{2}}}
\newcommand{\Ochh}{\ensuremath{\Oc h^{2}}}
\newcommand{\Ol}{\ensuremath{\Omega _{\Lambda}}}
\newcommand{\optdepth}{\tau}
\newcommand{\As}{A_\mathrm{s}}
\newcommand{\Asz}{A_\mathrm{SZ}}
\newcommand{\Ap}{A_\mathrm{p}}
\newcommand{\Cl}{C_{\ell}}
\newcommand{\Dl}{\mathcal{D}_{\ell}}
\newcommand{\VEV}{\PHI_{0}}
\newcommand{\PHI}{\phi}
\newcommand{\conj}{^*}
\newcommand{\FT}[1]{\tilde{#1}}
\newcommand{\vect}[1]{\mathbf{#1}}
\newcommand{\diff}{\mathrm{d}}
\newcommand{\Sr}{^{\mathrm{S}}}
\newcommand{\Vr}[1]{\!\!\stackrel{\scriptstyle{\mathrm{V}}}{_{\!\!#1}}}
\newcommand{\Tr}[1]{\!\!\stackrel{\scriptstyle{\mathrm{T}}}{_{\!#1}}}
\newcommand{\MM}{\mathcal{M}}

\newcommand{\unit}[1]{\;\mathrm{#1}}
\newcommand{\Eq}[1]{Eq. (\ref{eqn:#1})}
\newcommand{\Eqnb}[1]{Eq. \ref{eqn:#1}}
\newcommand{\Fig}[1]{Fig. \ref{fig:#1}}
\newcommand{\Sec}[1]{Sec. \ref{sec:#1}}
\newcommand{\Table}[1]{Table \ref{tab:#1}}

\newcommand{\PLSZ}{PL$_{\rm SZ}$}
\newcommand{\HZSZ}{HZ$_{\rm SZ}$}


\title{Cosmic string parameter constraints and model analysis using small scale Cosmic Microwave Background data}

\newcommand{\addressSussex}{Department of Physics \& Astronomy, University of Sussex, Brighton, BN1 9QH, United Kingdom}

\author{Jon Urrestilla}
\email{jon.urrestilla@ehu.es}
\affiliation{Department of Theoretical Physics, University of the Basque Country UPV-EHU, 48040 Bilbao, Spain}
\affiliation{\addressSussex}

\author{Neil Bevis} 
\email{n.bevis@imperial.ac.uk}
\affiliation{Theoretical Physics, Blackett Laboratory, Imperial College, London, SW7 2BZ, United Kingdom}

\author{Mark Hindmarsh} 
\email{m.b.hindmarsh@sussex.ac.uk}
\affiliation{\addressSussex}

\author{Martin Kunz}
\email{martin.kunz@physics.unige.ch}
\affiliation{D\'epartement de Physique Th\'eorique and Center for Astroparticle Physics, Universit\'e de Gen\`eve, 24 quai Ansermet, CH--1211 Gen\`eve 4, Switzerland}
\affiliation{\addressSussex}
\affiliation{African Institute for Mathematical Sciences, 6 Melrose Rd, Muizenberg, 7945, Cape Town, South Africa}

\date{\today}

\begin{abstract}
We present a significant update of the constraints on the Abelian Higgs cosmic string tension by cosmic microwave background (CMB) data, enabled both by the use of new high-resolution CMB data from suborbital experiments as well as the latest results of the WMAP satellite, and by
improved predictions  for the impact of Abelian Higgs cosmic strings on the CMB power spectra. The new cosmic string spectra \cite{Bevis:2010gj} were improved especially for small angular scales, through the use of larger Abelian Higgs string simulations and careful extrapolation.
If Abelian Higgs strings are present then we find improved bounds on their contribution to the CMB anisotropies, $\fd^{\rm AH} < 0.095$, and on their tension, $G\mu_{\rm AH} < 0.57\times 10^{-6}$, both at 95\% confidence level using WMAP7 data; and  $\fd^{\rm AH} < 0.048$ and  $G\mu_{\rm AH} < 0.42\times 10^{-6}$ using all the CMB data. We also find that using all the CMB data, a scale invariant initial perturbation spectrum, $n_s=1$, is now disfavoured at $2.4\sigma$ even if strings are present. A Bayesian model selection analysis no longer indicates a preference for strings.
\end{abstract}


\maketitle


\section{Introduction}
\label{sec:intro}

Observations of the cosmic microwave background (CMB) radiation \cite{WMAP7-C:Komatsu:2010fb,ACBAR=Reichardt:2008ay,QUAD=Friedman:2009dt,Dunkley:2010ge} and the large-scale distribution of galaxies \cite{Percival:2009xn} are consistent with the inflationary paradigm (see e.g.~\cite{Lyth:2009zz}) in which cosmic structure was seeded in the very early stages of the universe. Attempts at reconciling observations with a late-time generation of the perturbations have not proven very successful and appear to require a primordial mechanism like inflation \cite{Spergel:1997vq,Durrer:2001xu,Scodeller:2009iu}. Attention is now focused on the search for possible clues on the detailed physical processes that created the initial perturbations, through tests for departures from the standard cosmological model, which is characterised by Gaussian adiabatic density perturbations with a power-law power spectrum. An important 
clue would be the detection of phase transition remnants like topological defects, especially strings \cite{Vilenkin:1994book, Hindmarsh:1994re, Sakellariadou:2006qs, Copeland:2009ga,Copeland:2011dx,Hindmarsh:2011qj}.
A large class of physically-motivated inflation models predict the existence of cosmic strings, including hybrid inflation \cite{Shafi:1984tt,Yokoyama:1988zza,Dvali:1994ms,Garbrecht:2006az,Basboll:2011mh}, brane inflation in the context of string theory 
\cite{Sarangi:2002yt,Jones:2003da} and models rooted in supersymmetry and Grand Unified Theory (GUT) \cite{Jeannerot:2003qv}. Previous analyses have shown that cosmological datasets leave room for significant effects from cosmic strings  \cite{Bevis:2007gh, Battye:2006pk, Fraisse:2006xc, Wyman:2005tu}.

Data from the WMAP satellite, when interpreted via the standard $\Lambda$CDM model, suggests that the power spectrum of the primordial density perturbations obeys a power-law with index $\ns=0.963\pm0.015$ \cite{Dunkley:2008ie}. 
This appears to put some pressure on hybrid inflation models for which $\ns\simeq0.98$, and is significantly different from the exactly scale invariant Harrison-Zel'dovich spectrum ($\ns = 1$).
However, previous work \cite{Battye:2006pk,Bevis:2007gh} demonstrated that when cosmic strings are included the required value of $\ns$ can rise significantly, with the data then available implying $\ns=1.00\pm0.03$ with Abelian Higgs strings \cite{Bevis:2007gh}. So long as an inflation model can provide the appropriate relationship between $\ns$ and the string tension, the data can instead be seen to favour the model, with examples being the D3/D7 brane inflation model \cite{Haack:2008yb} and semi-shifted hybrid inflation in the extended supersymmetric Pati-Salam model \cite{Lazarides:2008nx}. Additionally, if other types of cosmic defect such as semilocal strings \cite{Urrestilla:2007sf} or cosmic textures \cite{Bevis:2004wk, Urrestilla:2007sf} are considered, then even larger $\ns$ values are possible, with those models additionally offering lower values of the string tension or symmetry-breaking scale \cite{Urrestilla:2007sf}.

Here we revisit the constraints upon Abelian Higgs strings, incorporating updates both on the theory and the data side:
We use more recent data, especially new observations on small angular scales ($\ell \gtrsim 2000$) where the string contribution has a greater relative importance \cite{Pogosian:2008am}.  The use of the small-scale data is enabled by a significant improvement in the CMB power spectrum predictions from Abelian Higgs string simulations \cite{Bevis:2010gj}, which benefits from improvements in computing power to achieve greater accuracy, and a better understanding of the generation of perturbations to extend the spectra to a multipole of $\ell = 4000$.

There has been other recent work on cosmic string constraints \cite{Battye:2010xz,Battye:2010hg,Dunkley:2010ge,Foreman:2011uj}, with different string models and datasets.  In \cite{Battye:2010xz} CMB data was taken from WMAP5 \cite{Dunkley:2008ie}, ACBAR \cite{ACBAR=Reichardt:2008ay}, BOOMERANG \cite{BOOMERANG=Jones:2005yb}, CBI \cite{CBI=Sievers:2009ah} and QUAD \cite{QUAD=Pryke:2008xp} experiments, and combined with galaxy power spectrum data from the SDSS Luminous Red Galaxy sample \cite{SDSS-LRG=Tegmark:2006az}.  The data was fitted with the Unconnected Segment Model (USM) of cosmic string power spectra 
\cite{Vincent:1996qr,Albrecht:1997mz,Pogosian:1999np}, with parameters chosen to resemble Nambu-Goto strings (USM-NG) or an earlier Abelian Higgs power spectrum (USM-AH) \cite{Bevis:2006mj}. The baryon density was constrained with a prior derived from estimates of the primordial Deuterium abundance in damped Lyman-$\alpha$ systems \cite{Pettini:2008mq}. In \cite{Dunkley:2010ge} CMB data was taken from WMAP7 \cite{WMAP7-C:Komatsu:2010fb} and ACT \cite{ACT-PS=Das:2010ga}, and fitted with the USM-NG cosmic string model, with priors on the Hubble parameter and BAO distance scales.  
The other parameters of the USM model beside the string tension can also be fitted to the data, rather than chosen to resemble numerical simulations \cite{Foreman:2011uj}.  The result is to weaken the constraints on the string tension. The apparent discrepancies between these works (including the present one) come both from the various datasets used and from the different input templates used for the string spectrum. The AH and the USM-NG cosmic string power spectra differ in shape, and more importantly in normalisation, leading to constraints on the cosmic string tension which are a factor of about 2 stronger for the the USM-NG model when fitted to the same datasets \cite{Battye:2010xz}.

In this paper, we compare our latest Abelian Higgs string CMB anisotropy power spectra, exhibited in Fig.~16 of \cite{Bevis:2010gj}, to CMB temperature and polarization measurements, using 7-year WMAP data \cite{WMAP7-C:Komatsu:2010fb} augmented by small-scale part-sky data from ACBAR \cite{ACBAR=Reichardt:2008ay} and QUAD \cite{QUAD=Friedman:2009dt}, and ACT \cite{Dunkley:2010ge}.  Due to the possibility that the presence of cosmic strings could introduce a scale dependent bias and induce early non-linear evolution in the matter power spectrum, we prefer to be conservative and do not use the SDSS galaxy power spectrum data (see also section~\ref{paramconst}).

We present new constraints for the Abelian Higgs model on the parameter $G\mu$, which is the product of Newton's constant $G$ and the string energy per unit length $\mu$, and the related parameter $\fd$, which is the fractional power in the temperature power spectrum due to cosmic strings at $\ell=10$. We also discuss the implication of string presence upon the index $\ns$, re-assessing the preference that we found for a 6-parameter cosmological model with Harrison-Zel'dovich perturbations and strings in \cite{Bevis:2007gh}. We further consider the effect of adding data from measurements of the Hubble parameter ($H_0$) \cite{Riess:2009pu}, baryon acoustic oscillations (BAO) \cite{Percival:2009xn} and the primordial deuterium abundance (BBN) \cite{Pettini:2008mq} as priors.  
We show that using CMB data alone, quite strong constraints can be derived without the uncertainty stemming from the use of other datasets, where the connection to the primordial perturbations is less easy to calculate.


\section{Data and Methodology}
\label{sec:MCMC}
Following the methodology used in  \cite{Bevis:2004wk} we add the defect  CMB angular power spectra and the primordial CMB anisotropy angular power spectra
(both temperature and polarisation spectra), and then explore the likelihood of these  combined $\Cl$ with the help
of the Markov chain Monte-Carlo (MCMC) method, using a modified version of CosmoMC \cite{Lewis:2002ah}.
While we re-calculate the inflationary contribution 
$\Cl^{\mathrm{inf}}(h,\ns,...)$ 
at each point in parameter space, we calculate the string contribution $\Cl^{\mathrm{str}}$ for a single set of cosmological parameters only. 
The normalization of string contribution to power spectrum is proportional to $(G\mu)^{2}$, where $\mu$ is the string mass per unit length and $G$ is Newton's constant, so that
\begin{eqnarray}
& \!\!\!\! \Cl(h,\ns,G\mu,...) & \\
& = \Cl^{\mathrm{inf}}(h,\ns,...)&
+ \left(\frac{G\mu}{G\mu_0}\right)^{2} \Cl^{\mathrm{str}}(h_0,G\mu_0,...),\nonumber
\end{eqnarray}
where $_0$ denotes the values of the cosmological parameters at which the string contribution was calculated: $h = 0.72$, 
$\Obhh = 0.214$, and
$\Omega_\Lambda = 0.75$, and an optical depth to last-scattering
of $\optdepth = 0.1$. 
This approximate treatment of the string component significantly reduces the computational cost of the MCMC procedure and is justified as the allowed changes in cosmology affect the string component at a level below the observational uncertainties.

We consider models both with and without strings and compare them to a number of datasets. To ease discussing them, we shall denote the standard 6-parameter cosmological model, with the power spectrum of primordial curvature perturbations obeying a power-law in $k$, as model PL. The parameters are: the index of this power-law $\ns$, its amplitude at $k=0.05 \unit{Mpc^{-1}}$ $\As$, the Hubble parameter $100h\unit{kms^{-1}Mpc^{-1}}$, the physical baryonic and cold dark matter density parameters $\Obhh$ and $\Ochh$, the optical depth to last scattering $\optdepth$. As standard in CosmoMC, we replace $h$ with $\theta$ which is approximately 100 times the ratio of the sound horizon at last scattering to the angular diameter distance. As observations extend to smaller angular scales it is now common to include a seventh parameter  
$\Asz$, the amplitude of the Komatsu and Seljak \cite{Komatsu:2002wc} template for the Sunyaev-Zel'dovich (SZ) effect, and we denote this model \PLSZ. 
Throughout our analysis we assume flat space, as would be yielded to a very good approximation by the relevant inflation models. Special cases of PL and \PLSZ\ are obtained by setting $\ns=1$, the Harrison-Zel'dovich spectrum, giving ``nested" models HZ and \HZSZ\ with 5 and 6 parameters respectively.

To models \PLSZ\ and \HZSZ\ we add the contributions to the CMB power temperature and polarization spectra from Abelian Higgs cosmic strings (AH) with free normalization $(G\mu_{\rm AH})^2$, which we label as \PLSZ+AH and \HZSZ+AH respectively. We then compare the above mentioned models  to seven-year WMAP data \cite{WMAP7-C:Komatsu:2010fb} ($\ell\lesssim1000$), and study the effect of adding ACBAR \cite{ACBAR=Reichardt:2008ay} ($300\lesssim\ell\lesssim3000$) and QUAD data \cite{QUAD=Friedman:2009dt} ($2000\lesssim\ell\lesssim3000$), and finally ACT data \cite{Dunkley:2010ge} ($\ell\lesssim10000$). 

\begin{figure}
\resizebox{\columnwidth}{!}{\includegraphics{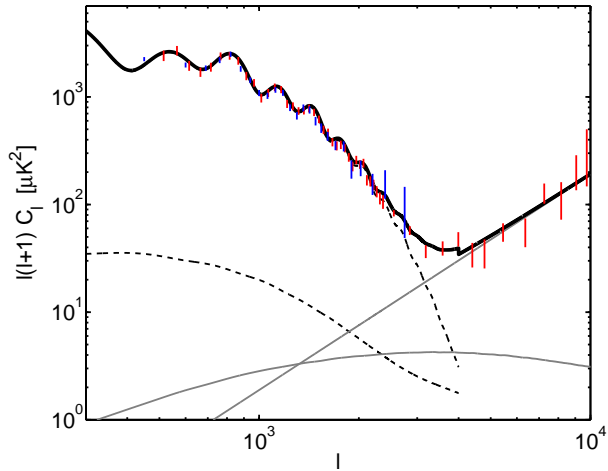}}
\caption{\label{fig:cl} CMB data and predictions for the temperature angular power spectrum from a model with strings.Parameters are those from the final column in Table I.
The thick black line corresponds to the total power spectrum, the upper dashed black line corresponds to the inflationary component, the lower dashed black line  corresponds to the string component, the humped grey line  corresponds to the SZ contribution (from the ACT 148 GHz template)  and the power-law grey line corresponds  to the Poisson point source model. ACT 148 GHz and ACBAR data are represented by red and blue bars respectively.
The small glitch at $\ell=4000$ is caused by the zeroing of the inflation and string components beyond this value. The result will be a slight overestimation of $\Ap$ and therefore potentially a very slight underestimation of strings.} 
\end{figure}

In order to fit ACT data, we also need to account for  foreground point sources, which we model as Poisson noise with a normalisation $\Ap$ \cite{Dunkley:2010ge}. As the point sources are the dominant contribution for $\ell > 4000$ we cut off the inflation and string power spectra for $\ell > 4000$. We show in Fig.~\ref{fig:cl} the data sets and the $\Cl$'s used.

We then investigate the effect of combining the CMB data with constraints from other measurements, which we impose as priors on the MCMC. 
We choose three of the most commonly used measurements: the Hubble parameter ($H_0$) \cite{Riess:2009pu}, the distance scales set by measurements of the Baryon Acoustic Oscillation (BAO) \cite{Percival:2009xn} and the baryon density inferred via standard Big Bang Nucleosynthesis from measurements of the primordial deuterium abundance (BBN) \cite{Pettini:2008mq}. 

\section{Parameter constraints}
\label{paramconst}
\subsection{CMB data only}

We initially consider the comparison of the models \PLSZ\ and \PLSZ+AH against (only)  CMB data, with results shown in \Table{MCMC}. Previously, when we performed a similar analysis to the three-year WMAP data we found a preference for a non-zero contribution from cosmic strings and an almost exactly scale-invariant power spectrum: $\ns=1.00\pm0.03$ \cite{Bevis:2007gh}.  We found that with WMAP7 the tendency remains, and both $\ns$ 
and $\Obhh$ tend to higher values when a string contribution is added. 

This increase in $\ns$ upon the inclusion of cosmic defects is due to a parameter degeneracy noted in Ref.~\cite{Bevis:2004wk}. When a contribution from cosmic defects is added to the adiabatic scalar component seeded by inflation, it is clear that $\As$ must be reduced if the fit to the WMAP data is to be maintained. However, the inflationary contribution is most sensitive to $\As$ at the first peak and therefore this peak is excessively lowered. This can be countered by an increase in $\Obhh$, which lifts the first peak while lowering the second. Additionally $\ns$ can be raised to lift the second peak relative to the first. Also, an increase in $h$ helps reverse the shift of the first peak to smaller scales which is caused by the latter, and also the shape of the string component.

\begin{table*}
\begin{tabular}{|c||c|c||c|c||c|c|}
\hline
Data &  \multicolumn{2}{c||}{WMAP7} &\multicolumn{2}{c||}{+ACBAR+QUAD}&\multicolumn{2}{c|}{ +ACT}\\
\hline
Model   & \PLSZ    & \PLSZ + AH & \PLSZ    & \PLSZ+ AH & \PLSZ   & \PLSZ+ AH\\
\hline
$\Obhh$ & $0.0224(6)$&$0.0235(10)$&$0.0225(5)$&$0.0231(6)$ & $0.0223(5)$&$0.0226(5)$\\
$\Ochh$ & $0.111(5)$&$0.101(6)$&$0.112(5) $&$0.109(5)$ &$0.113(5)$&$0.111(5)$\\
$\theta$& $1.039(3)$&$1.042(3)$&$1.041(2) $&$1.042(2)$ &$1.041(2)$&$1.042(2)$\\
$\optdepth$& $0.085(15)$&$0.094(17)$&$0.088(15)$&$0.086(15)$ &$0.087(14)$&$0.085(13)$\\
$\ns$   & $0.967(13)$&$0.985(20)$&$0.969(13)$&$0.975(13)$ &$0.965(12)$&$0.969(13)$\\
$\ln(10^{10}\As)$&$3.07(4)$&$3.07(5)$&$3.08(3) $&$3.05(3) $ &$3.09(3)$&$3.07(3)$\\
$\Asz$  &  $1.1(6)$&$0.9(6)$&$1.1(5)   $&$0.9(6) $ & $0.5(3)$ & $0.4(3)$\\
 $A_{p}$ &-&-&-&-&$16(2)$&$17(2)$\\
$10^{12}(G\mu)^2$& $-$&$0.13(10)$& -         &$0.11(7)$ & -&$0.07(5)$\\
\hline
$h$     & $0.708(23)$&$0.735(38)$&$0.712(23)  $&$0.731(25) $ & $0.703(23)$&$0.717(25)$\\
$\fd$     & $-$&$0.035(30)$&$-  $&$0.031(20)$ & - &$0.020(14)$\\
\hline
Likelihood&  $3734.044$&$3733.997$ &$3807.694$ &$3807.321$&$3823.176$&$3823.018$\\
\hline
\end{tabular}
\caption{\label{tab:MCMC}Marginalized likelihood constraints on model parameters. The basis parameters, upon which we assume flat priors, are listed above the horizontal line, with other derived parameters listed below this line. The notation is such that  $0.0224(6)\equiv0.0224\pm0.0006$, {\it i.e.}, it indicates an expectation value of $0.0224$ and a standard deviation uncertainty of $6$ in the final digit. 
In the first two columns, the data used are those from WMAP7. The next two columns use ACBAR and QUAD, as well as WMAP7. The last two columns use all the data from WMAP7, ACBAR, QUAD and ACT. Since these last two columns use ACT, we also include a $A_p$ to take into account Poisson point sources.}
\end{table*}

\begin{table*}
\begin{tabular}{|c||c||c||c||c||c||c||c||}
\hline
Data & $H_0$&BAO &BAO+$H_0$&BBN&BBN+$H_0$&BAO+BBN&BAO+BBN+$H_0$\\
\hline
$\Obhh$ &$0.0228(5)$& $0.0224(4)$&$0.0225(4)$&$0.0226(5)$&$0.0227(5)$&$0.0224(5)$&$0.0225(5)$ \\
$\Ochh$ &$0.110(3)$& $0.114(3)$&$0.113(3)$&$0.111(5) $&$0.110(4)$&$0.114(3) $&$0.113(3)$ \\
$\theta$& $1.042(2)$&$1.041(2)$&$1.041(2)$&$1.041(2) $&$1.042(2)$&$1.041(2) $&$1.041(2)$ \\
$\optdepth$&$0.087(14)$& $0.083(13)$&$0.085(14)$&$0.088(15)$&$0.090(15)$&$0.085(14)$&$0.086(14)$ \\
$\ns$   & $0.972(11)$&$0.964(11)$&$0.966(10)$&$0.968(12)$&$0.972(12)$&$0.963(11)$&$0.966(11)$ \\
$\ln(10^{10}\As)$&$3.06(3)$&$3.07(3)$&$3.07(3)$&$3.07(3) $&$3.07(3) $&$3.08(3) $&$3.08(3) $ \\
$\Asz$  & $0.4(3)$ &$0.4(3)$&$0.90(6)$&$0.4(3)   $ &$0.4(3) $&$0.4(3)   $ &$0.4(3) $\\
 $A_{p}$ &17(2)&17(2)&17(2)&17(2)&17(2)&17(2)&17(2)\\
$10^{12}(G\mu)^2$&$0.08(5)$ &$0.06(4)$&$0.07(5)$& 0.07(5)        &$0.08(5)$& 0.06(4)        &$0.06(5)$ \\
\hline
$h$     & $0.725(21)$&$0.701(15)$&$0.707(14)$&$0.714(24)  $&$0.723(20) $ &$0.700(15)  $&$0.706(14) $ \\
$\fd$     & $0.022(15)$&$0.016(12)$&$0.017(12)$&$0.019(13) $&$0.020(14)$&$0.016(12) $&$0.017(12)$ \\
\hline
\end{tabular}
\caption{\label{tab:MC2}Marginalized likelihood constraints on model parameters, as in Table~\ref{tab:MCMC}. The notation is such that $0.0228(5)\equiv0.0228\pm0.0005$, {\it i.e.}, it indicates an expectation value of $0.0228$ and a standard deviation uncertainty of $5$ in the final digit. 
In this case we fit \PLSZ+AH with all the CMB data (WMAP7, ACBAR, QUAD, ACT), with different priors. The different priors include (combinations of) priors  coming from  determinations of the Hubble parameter ($H_0$), Baryon Acoustic Oscillations (BAO) and Big Bang Nucleosynthesis (BBN), as explained in Section \ref{s:NonCMBDat}. }
\end{table*}

However, the increase in $\ns$ is not as high as with WMAP3, and the exactly scale-invariant power spectrum is not as favoured with the newer WMAP7. The mean $\ns$ value is reduced to $\ns=0.985\pm0.020$, and therefore there is now a slight preference for $\ns<1$, even when local cosmic strings are included. Hence inflation models that yield $\ns$ values greater than $1$ cannot be as easily aided by the inclusion of cosmic strings.

If we now add ACBAR and QUAD at high $\ell$, then we find additional constraints upon the string component, reducing its mean amplitude, as well as a  lower mean value for $\ns$.  Finally, further adding ACT data, the constraints on the string component get even more stringent, and the value of $\ns$ is now hardly shifted upwards when we allow for a string contribution. Use of the ACT data however necessitates the addition of SZ and point-source high-$\ell$ templates. Here we only quote results when allowing for the presence of unclustered point sources, which give rise to an extra contribution with constant $\Cl$. This population is detected at high significance ($A_p\approx 17\pm 2$, see tables \ref{tab:MCMC} and \ref{tab:MC2}).  We also experimented with adding a clustered point-source population \cite{Dunkley:2010ge}, finding that the amplitude of the template power spectrum was $A_c \approx3\pm3$.  As the result is consistent with zero, we did not include clustered point sources in our main results.

Table~\ref{gmu} shows a comparison of 95\% upper bounds on the string tension obtained by different data sets and different ways of modelling the string spectra. To date, the power spectra have been calculated using classical field theory simulations of the Abelian Higgs model \cite{Bevis:2006mj,Bevis:2010gj} or by the Unconnected Segment Model (USM), which represents the evolving string network with moving sticks with mass and tension. The sticks have random positions and velocities, and disappear so that their overall energy density scales as $t^{-2}$.  There is enough freedom in the USM to model the important scalar and vector components of the Abelian Higgs power spectra (USM-AH), and to make a prediction for the power spectra Nambu-Goto simulations would produce based on their mean square velocity and density (USM-NG). Note that there is currently no calculation of the CMB power spectra using NG simulations and the Unequal Time Correlator method of \cite{Bevis:2010gj}.

In general, we see that the WMAP7 data puts tighter constraints on the string contribution, mostly due to its smaller errors on the data points. The extra high-$\ell$ CMB experiments further push down the string contributions, where strings are expected to be more important. However, even though the constraints are tighter,  B-mode experiments such as CMBpol will be able to detect string components that are much lower than the bounds obtained here \cite{Mukherjee:2010ve}.

Note that the USM gives tighter constraints when modelling Nambu-Goto strings. This is mainly due to two features of NG simulations: 
a higher string density because of the absence of the decay channel into massive radiation \cite{Hindmarsh:2008dw}, and the persistence of the higher radiation era string density well into the matter era \cite{Battye:2010hg}.

\begin{table*}
\begin{tabular}{|c|c|c|c|}
\hline
Model & Data set &$10^6\, G\mu$ (95\%) & $\fd$ (95\%) \\
\hline
AH ~\cite{Bevis:2007gh}&WMAP3+BOOMERANG+CBI+ACBAR+VSA&0.7 & 0.11 \\
AH (this work)&WMAP7&0.57 & 0.095 \\
AH (this work)&WMAP7 + ACBAR + QUAD + ACT& 0.42 & 0.048  \\
USM-AH~\cite{Battye:2010xz}&WMAP5&0.68 & 0.11 \\
USM-NG ~\cite{Battye:2010xz}&WMAP5&0.28 & 0.054 \\
USM-NG~\cite{Dunkley:2010ge}&WMAP7+ACT& 0.16  & \\
\hline
\end{tabular}
\caption{\label{gmu}CMB constraints on string tension obtained by different string modeling, and different data sets.  The table shows 95\% upper bounds on $G\mu$. The different models shown are the Abelian Higgs model (AH) and the Unconnected String Segment model (USM), with parameters chosen to resemble the Nambu-Goto (NG) or Abelian Higgs (AH) spectra. The difference in the values of $G\mu$ and $\fd$ are due to both the different string templates used and the different data sets used.}
\end{table*}

\begin{figure}
\resizebox{\columnwidth}{!}{\includegraphics{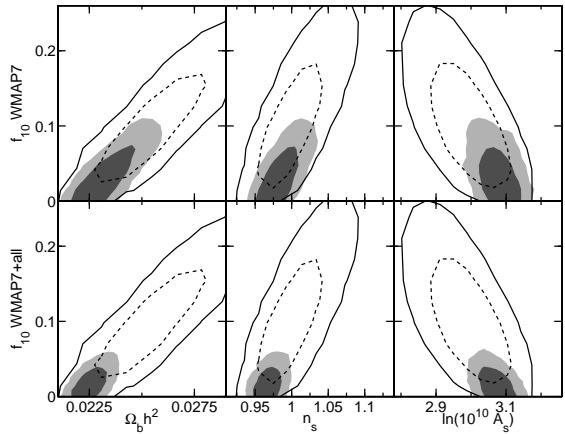}}
\caption{\label{comp} Marginalised 2-D likelihood contours for different data sets. The unshaded regions in both the top and bottom panel correspond to the 1-$\sigma$ and 2-$\sigma$ contours obtained in \cite{Bevis:2007gh} when contrasted with WMAP3 data \cite{WMAP3-T=Hinshaw:2006ia}. The shaded regions correspond to the 1-$\sigma$ and 2-$\sigma$ contours  obtained in the present work. The top panel shows the contours for WMAP7 data, and the bottom panel shows the contours for WMPA7+AQBAR+QUAD+ACT.} 
\end{figure}

Figure~\ref{comp} shows marginalised 2-D likelihoods of some of the parameters in the models, and depicts how the constraints have shrunk by taking into account the new data. The contours obtained with WMAP3 data were much wider, and it can readily be seen that $\ns=1$ was close to the mean value. The degeneracies among the parameters are also apparent. The inclusion of WMAP7 data made the contours smaller by more than a factor of two, leaving less space for strings, and pushing $\ns$ away from scale invariance. By including the rest of the CMB data mentioned above, the contours shrink an extra factor of two. 

\subsection{Adding non-CMB data}
\label{s:NonCMBDat}

In this section we investigate the effect of combining the CMB data with constraints from other measurements, the Hubble parameter ($H_0$) \cite{Riess:2009pu}, the distance scales set by measurements of the Baryon Acoustic Oscillation (BAO) \cite{Percival:2009xn} and the baryon density inferred via standard Big Bang Nucleosynthesis from measurements of the primordial deuterium abundance (BBN) \cite{Pettini:2008mq}. We include the data as priors on the MCMC.

We see that the most recent determination of $H_0$ of $74.2\pm3.6\; \mathrm{km}\,\mathrm{s}^{-1} \, \mathrm{Mpc}^{-1}$ from a combination of Cepheids and SN-Ia data \cite{Riess:2009pu} is slightly higher than the CMB mean value of $70.3\pm2.3\; \mathrm{km}\,\mathrm{s}^{-1} \, \mathrm{Mpc}^{-1}$ under the \PLSZ\ model. Imposing this prior on $H_0$ for the \PLSZ+AH model  allows for a slightly higher value of $\fd$, but the effect is negligible.

Distance measurements with the help of the baryon acoustic oscillations (BAO) are supposed to be very reliable as they are based on a physical mechanism closely related to the one that created the acoustic oscillations in the CMB. However, the resulting oscillatory feature in the galaxy power spectrum is on intermediate scales where the matter power spectrum $P(k)$ is on the verge of deviating from the linear prediction already in the standard cosmological model without strings. In addition, measurements of the galaxy power spectrum or correlation function are always subject to galaxy bias. The reliability of the BAO measurement has been tested with large N-body simulations in the absence of topological defects \cite{Crocce:2007dt} but if cosmic strings are present then one would need to check whether they are able to affect either the galaxy bias or the scale on which perturbations turn non-linear. This requires the inclusion of string perturbations in N-body simulations \cite{Obradovic:2011mt}. In any case the effect of adding the BAO data is to shift mean values by less than half a standard deviation.

The noted increase in $\Obhh$ for PL+AH is of interest since model PL already yields values of this parameter which are larger than those measured by alternative means. Big bang nucleosynthesis (BBN) calculations link the primordial abundance of the light isotopes to $\Obhh$ but it is only possible to measure such abunda<nces in astronomical objects at more recent times. While $\Obhh$ measures using different isotopes do not wholly agree, deuterium abundances in high redshift gas clouds ($z\approx3$) can be measured via absorption lines in quasar spectra, yielding a result of $\Obhh=0.0213\pm0.0010$ \cite{Pettini:2008mq}. Assuming a Gaussian uncertainty, this can be included into our analysis, yielding the results in \Table{MC2}. However, given that there are only 7 measurements, with abundances corresponding to $\Obhh$ in the range 0.017 to 0.030, the BBN constraint should be treated with caution.
The effect on model \PLSZ~is a small reduction (less than half a standard deviation) in $\Obhh$ via a degeneracy similar to that noted for the defect case above.  The effect on the parameters $\ns$ and $\fd$ of the model \PLSZ+AH is negligible, showing that CMB data are now good enough to render the BBN constraint unnecessary.

For these reasons we still consider the CMB-only constraints as the most reliable results, which we therefore use as the main figures quoted in this paper. In any case, the addition of the small-scale CMB data is  as powerful as the addition of the non-CMB data considered here.

\section{Model selection}

We now compare models via maximum likelihood values $\mathcal{L}_{\max}$ and also via Bayesian evidence ratios, which yields a more complete analysis of the freedom within a model (see, for example, \cite{Liddle:2006tc}). We determine maximum likelihood values via low temperature MCMC chains and then express results via $\Delta\chi^2_\mathrm{eff}=-2\Delta\ln(\mathcal{L}_{\max})$ between two models. We determine evidence ratios using the Savage-Dickey density ratio \cite{Kunz:2006mc,Trotta:2005ar}, which can be used with nested models. This yields the Bayes factor between an $n+1$ parameter model $\mathcal{M}_{n+1}$ (such as \PLSZ+AH) and an $n$ parameter subset of that model $\mathcal{M}_{n}$ (such as \PLSZ), via:
\begin{equation}
B_{n+1/n} 
= 
\frac{P(d|\mathcal{M}_{n+1})}{P(d|\mathcal{M}_{n})} 
= 
\frac{\pi(x=x_*|d,\mathcal{M}_{n+1})}{P(x=x_*|d,\mathcal{M}_{n+1})},
\end{equation}
where $x$ is the $(n+1)$th parameter and $x_*$ its fixed value in the $n$ parameter model. That is, the complex calculation of evidences is reduced to determining the value of the (independent) prior upon $x$ in model $\mathcal{M}_{n+1}$, $\pi(x=x_*|d,\mathcal{M}_{n+1})$, and the marginalized posterior probability density for $x$ at $x_*$ (given the data $d$ and model $\mathcal{M}_{n+1}$) $P(x=x_*|d,\mathcal{M}_{n+1})$. The former is trivial for the flat priors that we assume here but highlights the dependence of the results upon prior choice. The latter, on the other hand, can be found via binning the MCMC results according to their $x$ values. 

\begin{table}
\begin{tabular}{|l|c|c|}
\hline
Data     & \multicolumn{2}{c|}{CMB only}                 \\
\hline
Measure  & $\Delta\chi^2_\mathrm{eff}$ & $\ln B$       \\
\hline
\PLSZ    & 0                           & $0$ \\
\PLSZ+AH &       $-0.3$            & $-3.3\pm0.1$ \\ 
\HZSZ    &           $+7.8$            & $-1.0 \pm 0.3$ \\  
\HZSZ+AH &      $+4.7$            & $-3.2\pm0.2$ \\
\hline   
\end{tabular}
\caption{\label{ModelSelection}  $\Delta\chi^2$ and the natural logarithm of the Bayes factors ($\ln B$) for four different models, taking as the base model \PLSZ. The data used in these comparisons are the full CMB data used in this work, which means that ACT is also taken into account, and the models have  foregrounds modelled with Poisson point sources. Therefore, \PLSZ~is an 8 parameter model, with the usual 6 parameters plus $\Asz$ and $\Ap$.}
\end{table}

Our results for both statistics are shown in Table~\ref{ModelSelection} for the models employed above, based around adding an AH component to model \PLSZ. We now also include results for model \HZSZ\ (with and without an AH component). These models are nested into each other at points $\ns=1$ and $(G\mu)^2=0$. There are hence two flat priors whose range is relevant to our results and for these we take: $\ns=0.75\rightarrow1.25$ and $(G\mu)^2=0\rightarrow(2\times10^{-6})^2$. The latter is inspired by the approximate limit that we would obtain if we compared our string spectra to COBE data ($\ell\lesssim10$) \footnote{This is roughly equivalent near the nesting point $\fd=0$ to the prior of $\fd=0\rightarrow1$ that we employed in Ref. \cite{Bevis:2007gh}, although then the limits were simply mathematical constraints on a fractional quantity and near \fd=1 the two priors would noticeably differ.}. 

While in \cite{Urrestilla:2007sf} we found that the addition of an AH string component to the PL model lowered the $\chi^2$ by $3.5$ and lead to a comparable model probability for PL+AH (and even a somewhat higher model probability for HZ+AH) we now find that all alternative models are weakly to moderately disfavoured when compared to \PLSZ. There are two reasons for the change: Firstly the improvement of fit when adding AH to \PLSZ~has decreased, with \HZSZ+AH now disfavoured, and secondly the posterior parameter volume (i.e. the error bar on $(G\mu)^2$) has also decreased, which increases the ``Occams razor'' penalty for a fixed prior size. The change in goodness of fit (due to $n_s=1$ being disfavoured now even when strings are present) is especially important for the \HZSZ+AH model.

Given the relative model probabilities, we are able to take into account the model uncertainty in the
determination of $(G\mu)^2$ or $f_{10}$ by marginalising over the models (see e.g. \cite{Parkinson:2010zr}):
\begin{equation}
P(f_{10}|d) = \sum_\MM P(f_{10},\MM|d) = \sum_\MM P(f_{10}|d,\MM) P(\MM|d) .
\end{equation}
The sum over models here runs over the set \{\PLSZ, \HZSZ, \PLSZ+AH, \HZSZ+AH  \}.
If we take the prior probability of all models to be equal then we can replace the model probability $P(\MM|d)$ with $P(d|\MM)$, and since we determine the overall normalisation of the model-averaged posterior by requiring that the integral over $f_{10}$ be equal to 1 we can in addition replace $P(d|\MM)$ with Bayes factors relative to one of the models.

\begin{figure}
\resizebox{8cm}{!}{\includegraphics{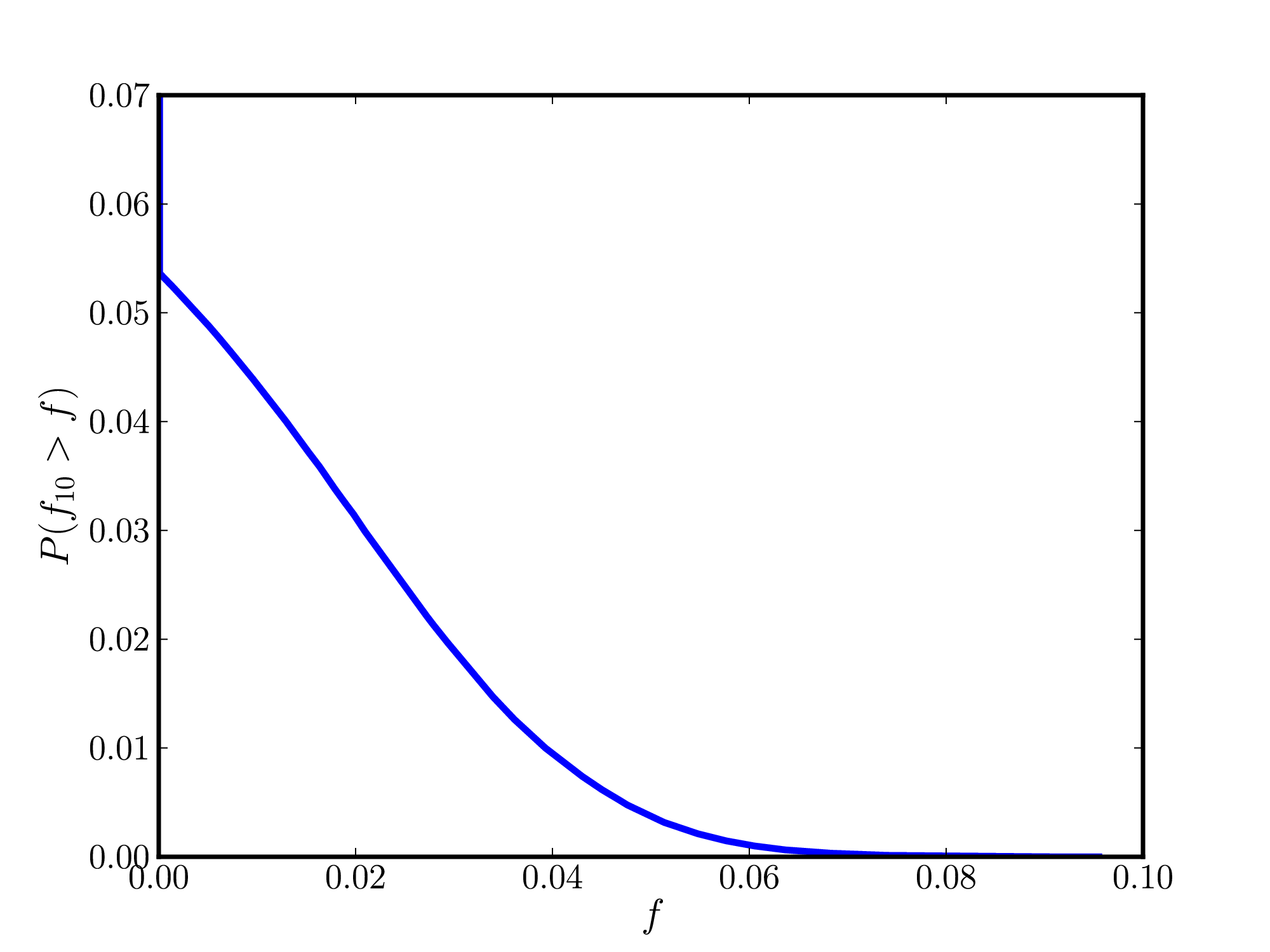}}
\caption{\label{fig:modave}Cumulative model-averaged 1D probability distribution for $\fd$ using the Bayes factors in table \ref{ModelSelection}. For the models and priors chosen, 94.6\% of the weight is in the ``no strings'' models \PLSZ~and \HZSZ.  From the choice of priors and models, this figure gives an idea of the constraints on generic scenarios where Abelian-Higgs strings are formed at an energy scale associated with inflation (see text for more details on priors and model choices).}
\end{figure}

Since for the two models without strings we have that $P(f_{10}) = \delta(f_{10})$, which is difficult to plot, in Fig.~\ref{fig:modave} we show the cumulative probability that $\fd$ is larger than a given value. Based on the Bayes factors in table \ref{ModelSelection}, we find that 94.6\% of the posterior volume is in the two ``no-string'' models \PLSZ~and \HZSZ. The use of the model-averaged posterior would therefore tighten the 95\% upper limit on strings significantly. However, this result is strongly dependent on the selection of models considered, and we prefer to quote upper limits on $\fd$ and $G\mu$ under the assumption that Abelian-Higgs strings are present and that the initial power spectrum is of a power-law form, i.e., the model \PLSZ+AH.

Note that the Bayes factors in Table \ref{ModelSelection} also depend strongly on the priors used, especially on the fact that we assume $G\mu$ to be uniformly distributed in the range $[0,2\times 10^{-6}]$. This prior appears appropriate for ``GUT strings'' associated with the energy scale of inflation. If we are interested in constraints on strings that were formed in a phase transition at an arbitrary energy scale, then a flat prior in $\ln G\mu$ would be better motivated. Such an alternative prior would change the result strongly, as the prior weight  is then equally distributed over the range of possible scales for $G\mu$ (a Jeffreys prior \cite{Jeffreys24091946})\footnote{Since the nesting point is $G\mu=0$ the Jeffreys prior would prevent the direct use of the Savage-Dickey density ratio, but see \cite{Mukherjee:2010ve} for a way to circumvent this problem.}. Such a prior, with a sensible choice of the $\ln G\mu$ lower bound, would place a large prior weight on very low $G\mu$ values where the posterior is flat. As the posterior is close to maximal for $f_{10}\rightarrow 0$ with current data it is clear that a strong preference for or against strings is impossible to obtain in both model selection and the above parameter results, resulting in roughly equal model probabilities for \PLSZ~and \PLSZ+AH. We refer the reader to the detailed discussion of the computational methods and the expected effects of a logarithmic prior on $G\mu$ in \cite{Mukherjee:2010ve}.


\section{Conclusions}

In this paper we use an improved determination of the Abelian Higgs cosmic string CMB power spectrum $\Cl$ up to $\ell = 4000$, combined with new CMB data that extends out to $\ell \approx 10000$ to update our constraints on the maximal contribution of strings to the CMB anisotropies. We find that $\fd^{\rm AH} < 0.048$ at 95\% confidence level, which translates into a bound on the Abelian Higgs string tension of $G\mu_{\rm AH} < 0.42$. Using only WMAP7 data, the bounds we obtain are $\fd^{\rm AH} < 0.095$ and $G\mu_{\rm AH} < 0.57$ at  95\% confidence level. These bounds are somewhat weaker than the ones derived by other groups using the unconnected segment model for Nambu-Goto strings, which is mostly due to a lower string density in the field theory simulations. Note that even though the constraints are tighter than previous ones, the Planck satellite is expected to improve them (to about $ \fd < 0.01$) \cite{Urrestilla:2008jv}, and a future dedicated  B-mode satellite will be able to detect string components that are much lower than the bounds obtained here \cite{Mukherjee:2010ve,Foreman:2011uj}.

While the WMAP 7-year data set already improves the constraints significantly compared to the 3-year data set,  the additional constraints from high-$\ell$ data (ACBAR, QUAD and especially ACT) help to break the degeneracy between the amplitude, $n_s$ and $\Omega_b h^2$ so that a Harrison-Zel'dovic spectrum with $n_s=1$ is now also disfavoured when allowing for the presence of strings. Because of this, we now obtain strong constraints with CMB data alone, and the addition of BAO, BBN and $H_0$ constraints is less important than in the past.

Previously Bayesian model selection slightly favoured HZ+AH over PL when using only CMB data \cite{Bevis:2007gh,Urrestilla:2007sf}. This has changed with the improved CMB data that no longer prefers a non-zero $\fd^{\rm AH}$ and also puts pressure on a Harrison-Zel'dovich initial power spectrum even with AH strings. For the choice of priors and models investigated here we find that models with AH strings are disfavoured. 

We expect that the strong constraints obtained in this work put pressure on generic models that form strings of the Abelian-Higgs type at energy scales associated with inflation, e.g., these constraints push hybrid inflation models to small couplings. We postpone the study of the constraints on specific implementations of  high energy physics based models of inflation that form defects to later work. Generically, in those models, the string tension is not a free parameter, since it is related to other measurable predictions. Having to allow for a low string tension will put pressure on other parameters to meet the measurements, and it will be interesting to investigate whether the interplay between the parameters  is able to render the model viable.


\begin{acknowledgments}
We thank Joanna Dunkley for discussion and for providing us with an early version of the ACT likelihood code, and Arttu Rajantie for useful discussions. We acknowledge support from the Science and Technology Facilities Council [grant numbers ST/G000743/1, ST/F002858/1, ST/I000976/1]
(NB, MH) and the Swiss National Science Foundation (MK).  J.U. acknowledges financial support from the Basque Government (IT-559-10), the Spanish Ministry (FPA2009-10612), and the Spanish Consolider-Ingenio 2010 Programme CPAN (CSD2007-00042). Numerical calculations were performed using the UK National Cosmology Supercomputer (supported by SGI/Intel, HEFCE, and STFC), the Imperial College HPC facilities, the University of Sussex HPC Archimedes cluster and the Andromeda cluster of the University of Geneva. 
\end{acknowledgments}

\bibliography{DATAarxivNotes}

\end{document}